\documentclass[sigconf]{acmart}

\copyrightyear{2026}
\acmYear{2026}
\setcopyright{cc}
\setcctype{by}
\acmConference[SIGIR '26] {Proceedings of the 49th International ACM SIGIR Conference on Research and Development in Information Retrieval}{July 20--24, 2026}{Melbourne, VIC, Australia.}
\acmBooktitle{Proceedings of the 49th International ACM SIGIR Conference on Research and Development in Information Retrieval (SIGIR '26), July 20--24, 2026, Melbourne, VIC, Australia}
\acmISBN{979-8-4007-2599-9/2026/07}
\acmDOI{10.1145/3805712.3809857}

\begin{document}

\title{Equip Pre-ranking with Target Attention by Residual Quantization}

\author{Yutong Li}
\authornote{Both authors contributed equally to this research.}
\email{mark.lyt@taobao.com}
\author{Yu Zhu}
\authornotemark[1]
\authornote{Corresponding author.}
\email{zy143829@taobao.com}
\affiliation{%
  \institution{Taobao \& Tmall Group of Alibaba}
  \city{Hangzhou}
  \country{China}
}

\author{Yichen Qiao}
\affiliation{%
  \institution{Shanghai Jiao Tong University}
  \city{Shanghai}
  \country{China}}
\email{yichen1@sjtu.edu.cn}

\author{Ziyu Guan}
\affiliation{%
 \institution{Xidian University}
 \city{Xi'an}
 \country{China}}
\email{zyguan@xidian.edu.cn}

\author{Lv Shao}
\affiliation{%
  \institution{Taobao \& Tmall Group of Alibaba}
  \city{Hangzhou}
  \country{China}}
\email{shaolv.sl@taobao.com}

\author{Tong Liu}
\affiliation{%
  \institution{Taobao \& Tmall Group of Alibaba}
  \city{Hangzhou}
  \country{China}}
\email{yingmu@taobao.com}

\author{Bo Zheng}
\affiliation{%
  \institution{Taobao \& Tmall Group of Alibaba}
  \city{Beijing}
  \country{China}}
\email{bozheng@alibaba-inc.com}

\renewcommand{\shortauthors}{Yutong Li et al.}

\begin{abstract}
The pre-ranking stage in industrial recommendation systems faces a fundamental conflict between efficiency and effectiveness. While powerful models like Target Attention (TA) excel at capturing complex feature interactions in the ranking stage, their high computational cost makes them infeasible for pre-ranking, which often relies on simplistic vector-product models. This disparity creates a significant performance bottleneck for the entire system. To bridge this gap, we propose TARQ, a novel pre-ranking framework. Inspired by generative models, TARQ's key innovation is to equip pre-ranking with an architecture approximate to TA by Residual Quantization. This allows us to bring the modeling power of TA into the latency-critical pre-ranking stage for the first time, establishing a new state-of-the-art trade-off between accuracy and efficiency. Extensive offline experiments and large-scale online A/B tests at Taobao demonstrate TARQ's significant improvements in ranking performance. Consequently, our model has been fully deployed in production, serving tens of millions of daily active users and yielding substantial business improvements. The code and data are available at https://github.com/zyody/tarq\_sigir2026.
\end{abstract}




\maketitle
\begin{figure}[h]
  \centering
  \includegraphics[width=\linewidth]{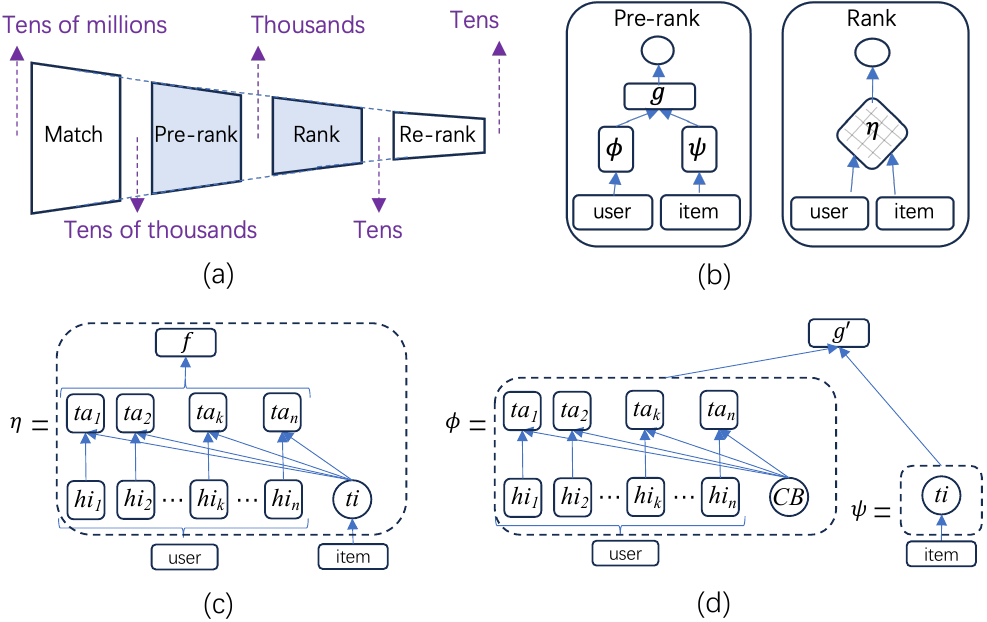}
  \caption{(a) Multi-stage architecture: matching, pre-ranking, ranking and re-ranking with scoring numbers. (b) Representation-focused models for pre-ranking and interaction-focused models for ranking. (c) The structure of TA. It dynamically re-weights a user's historical behaviors $hi_1,hi_2…hi_n$ based on their relevance to a specific target item $ti$. (d) The structure of our proposed TARQ, which equips pre-ranking with TA in a representation-focused manner. $CB$ means codebook.}
  \Description{A woman and a girl in white dresses sit in an open car.}
  \label{fig:multistage}
\vspace*{-25pt}
\end{figure}
\section{Introduction}
Modern industrial recommendation systems are tasked with retrieving a small subset of relevant items from a massive candidate pool. The de facto industry standard is to adopt a multi-stage cascaded architecture \cite{gallagher2019joint, liu2017cascade} (shown in Figure \ref{fig:multistage} (a)). The ranking stage typically leverages interaction-focused models \cite{guo2020deep} to capture complex feature interactions for accuracy, while simpler representation-focused models \cite{guo2020deep} are adopted in pre-ranking to ensure efficiency.

As illustrated in Figure \ref{fig:multistage} (b), representation-focused models prioritize efficiency by pre-computing item representations ($\psi$) offline. During online serving, they only need to compute the user representation ($\phi$) once and then perform efficient vector-product scoring ($g$) across a large candidate set of size $N_1$. The total online cost is thus $\phi + N_1 \times g$. Conversely, the ranking stage utilizes interaction-focused models that perform complex feature interaction computations ($\eta$) for each of the $N_2$ candidates, resulting in an online cost of $N_2 \times \eta$. Given that the cost of $g$ is orders of magnitude lower than $\eta$, pre-ranking can operate on a significantly larger candidate set (i.e., $N_1 \gg N_2$ ). The architectural simplicity of representation-focused pre-ranking models, optimized aggressively for efficiency, imposes a severe penalty on effectiveness. Target Attention (TA) has emerged as a cornerstone of modern interaction-focused ranking models \cite{zhou2018deep, zhou2019deep, feng2019deep, chen2019behavior, pi2020search, chen2021end, cao2022sampling, si2024twin, chang2023twin}. The structure is detailed in Figure \ref{fig:multistage} (c). However, despite its proven effectiveness, the prohibitive computational cost of TA has precluded its adoption in the latency-critical pre-ranking stage.

To bridge this modeling gap, we introduce \textbf{TARQ}, a novel pre-ranking framework that brings the power of TA into the pre-ranking stage. Our key innovation, inspired by recent advances in generative models, is to equip pre-ranking with an architecture approximate to \textbf{TA} by \textbf{R}esidual \textbf{Q}uantization. Specifically, as shown in Figure \ref{fig:multistage} (d), TARQ decouples the expensive user-target interaction. It first computes attention scores between the user's historical behaviors and a shared, quantized codebook (CB). Then the attention scores on $ti$ are efficiently assembled through a rapid index lookup operation $g'$, which adds only marginal computational overhead to the traditional representation-focused models. This allows TARQ to capture the fine-grained interaction patterns of TA while satisfying the strict efficiency constraints of pre-ranking, fundamentally improving the quality of the candidate set for downstream stages. In addition, different from previous \textit{Generative} Retrieval works \cite{rajput2023recommender, feng2022recommender, tay2022transformer, zheng2024adapting} which utilize Residual Quantization (RQ) in a sequence generation task, we repurpose RQ for approximating the TA mechanism within a traditional \textit{discriminative} \cite{wang2025scaling} pre-ranking framework, and further mitigate
the codebook collapse issue by the designed Codebook Alignment objective in our task. Our main contributions are summarized as follows:
\begin{itemize}
\item We propose TARQ, a novel pre-ranking framework that successfully incorporates the expressive power of TA into the latency-critical pre-ranking stage, establishing a new state-of-the-art trade-off between performance and efficiency.
\item The proposed Codebook Alignment technique effectively mitigates codebook collapse, boosting utilization from a baseline of 59\% to 98\% in our task. This highlights its potential as a general solution for this common issue in other quantization-based architectures.
\item We demonstrate the effectiveness and efficiency of TARQ through extensive offline evaluations and large-scale online A/B tests on the Taobao e-commerce platform, confirming its practical value in a real-world industrial setting.


\end{itemize}

\section{PROPOSED METHOD}

\begin{figure*}[h]
  \setlength{\abovecaptionskip}{0cm}
  \setlength{\belowcaptionskip}{0.5cm}
  \centering
  \includegraphics[width=0.8\linewidth]{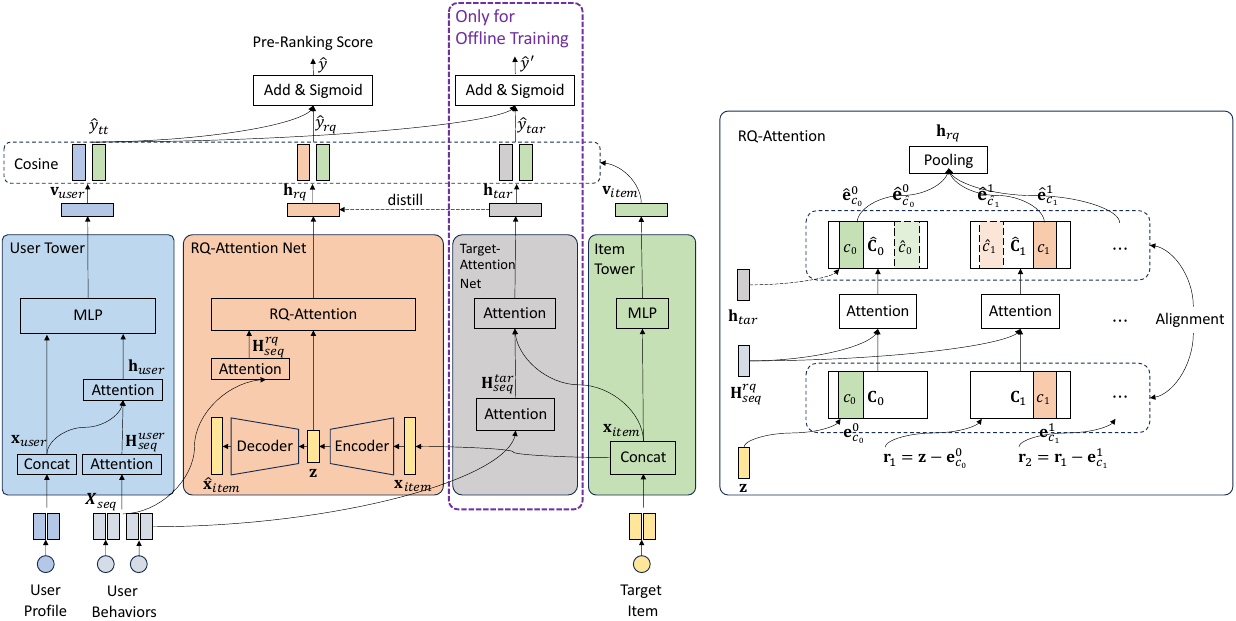}
  \caption{The overall architecture of TARQ}
  \Description{A woman and a girl in white dresses sit in an open car.}
  \label{fig:model}
\vspace*{-25pt}
\end{figure*}

As illustrated in Figure \ref{fig:model}, our proposed TARQ model adopts a teacher-student architecture comprising three core components: (1) a standard two-tower backbone; (2) an online student (the RQ-Attention Net), which efficiently approximates the TA mechanism at inference; and (3) an offline teacher (the Target-Attention Net), which generates high-fidelity interaction signals during training to supervise the student.

\subsection{Two-Tower Backbone}
\noindent{\bfseries User Tower.} 
The user tower generates a comprehensive user representation, $\mathbf{v}_{user}$, by integrating static profile features with dynamic behavioral data. Given the user profile embedding $\mathbf{x}_{user}$, and the historical item sequence $\mathbf{X}_{seq}=[\mathbf{x}^1_{i},\mathbf{x}^2_{i}, \ldots, \mathbf{x}^n_{i}]$, the process is as follows:
\begin{align}
\mathbf{H}_{seq}^{user}&={\rm MHA}(\mathbf{X}_{seq}, \mathbf{X}_{seq}, \mathbf{X}_{seq})  \label{eq:selfAtt},\\ 
\mathbf{h}_{user}&={\rm MHA}(\mathbf{x}_{user}, \mathbf{H}_{seq}^{user}, \mathbf{H}_{seq}^{user}) \label{eq:crossAtt},\\ 
\mathbf{v}_{user}&={\rm MLP}([\mathbf{x}_{user}; \mathbf{h}_{user}])\label{eq:userRep},
\end{align}
where MHA denotes the Multi-Head Attention operation \cite{vaswani2017attention}.\\
\noindent{\bfseries Item Tower.} Symmetrically, the item tower generates the item representation $\mathbf{v}_{item}$ by transforming the item's embedding $\mathbf{x}_{item}$ through its own MLP: 
\begin{align}
\mathbf{v}_{item}&={\rm MLP}(\mathbf{x}_{item}) \label{eq:itemRep}.
\end{align}

\noindent{\bfseries Prediction.} The final prediction score of this two-tower backbone, denoted $\hat y_{tt}$, is computed as the cosine similarity between them:
\begin{align}
\hat y_{tt}&={\rm cos}(\mathbf{v}_{user}, \mathbf{v}_{item})\label{eq:pred}.
\end{align}

\subsection{RQ-Attention Net}
The RQ-Attention Net aims to approximate the performance of TA. 

\noindent{\bfseries AutoEncoder.} Unlike the original RQ-VAE \cite{lee2022autoregressive}, we separate the reconstruction and quantization steps to reduce information loss during the process. The AutoEncoder is solely responsible for the reconstruction task. It takes target item features embedding $\mathbf{x}_{item}$ as input, which is passed through a DNN Encoder to obtain a latent variable $\mathbf{z}$. A DNN Decoder then reconstructs $\mathbf{z}$ into $\mathbf{\hat x}_{item}$. The reconstruction loss function is defined as follows:
\begin{align}
{\mathcal L}_{recon}&=||\mathbf{x}_{item}-\mathbf{\hat x}_{item}||_2^2\label{eq:reconLoss}.
\end{align}

\noindent{\bfseries Residual-Quantizer.} 
Residual-Quantizer is a multi-level technique that approximates a high-dimensional vector $\mathbf{z}$ by iteratively quantizing its residuals. Given $m$ codebooks $\{\mathbf{C}_l\}_{l=0}^{m-1}$, the process begins with the initial residual $\mathbf{r}_0=\mathbf{z}$. At each level $l$, the current residual $\mathbf{r}_l$ is quantized by finding the nearest entry $\mathbf{e}^l_{c_l}$ in the codebook $\mathbf{C}_l=\{\mathbf{e}_k^l\}_{k=1}^K$, where ${c_l}$ is the resulting semantic ID: $c_l={\rm argmin}_{i}||\mathbf{r}_l-\mathbf{e}^l_i||$. The residual for the next level is then updated: $\mathbf{r}_{l+1} = \mathbf{r}_l - \mathbf{e}^l_{c_l}$. This is repeated $m$ times, yielding a list of semantic IDs $\mathbf{s}=[c_0, c_1, \ldots, c_{m-1}]$ and their corresponding vectors $\mathbf{E}_{rq}=[\mathbf{e}^0_{c_0},\mathbf{e}^1_{c_1},\ldots,\mathbf{e}^{m-1}_{c_{m-1}}]$. 


The loss function of resudual-quantizer is:
\begin{align}
{\mathcal L}_{rq}=\sum_{l=0}^{m-1}||{\rm sg}[\mathbf{r}_l]-\mathbf{e}_{c_l}^l||_2^2\label{eq:rqLoss},
\end{align}
where ${\rm sg}[\cdot]$ is the stop-gradient operation.

\noindent{\bfseries RQ-Attention.} 
The RQ-Attention Net's core innovation is a \textit{personalized codebook mechanism} that efficiently approximates TA. The process unfolds in the following two key stages.\\
Codebook Personalization: Instead of using the static RQ codebooks, we dynamically adapt them for the current user by treating each codebook as a query set against the user's contextualized history:
\begin{align}
\mathbf{H}^{rq}_{seq}&={\rm MHA}(\mathbf{X}_{seq}, \mathbf{X}_{seq}, \mathbf{X}_{seq}) \label{eq:selfAtt2},\\ 
\mathbf{\hat C}_l=\{\mathbf{\hat e}_k^l\}_{k=1}^K&={\rm MHA}(\mathbf{C}_l,\mathbf{H}^{rq}_{seq},\mathbf{H}^{rq}_{seq})\label{eq:personalCB}.
\end{align}
\\
Approximate Interaction: Using the original semantic IDs $\mathbf{s}=[c_0, c_1, \ldots, c_{m-1}]$ derived from the target item, we look up the corresponding vectors $\mathbf{\hat E}_{rq}=[\mathbf{\hat e}^0_{c_0}, \mathbf{\hat e}^1_{c_1}, \ldots, \mathbf{\hat e}^{m-1}_{c_{m-1}}]$ from these newly personalized codebooks $\{\mathbf{\hat C}_l\}_{l=0}^{m-1}$. These personalized vectors are then fused (e.g., via pooling) to form the final approximate representation TA output, $\mathbf{h}_{rq}$. The relevance score $\hat y_{rq}$ is then computed.
\begin{align}
\mathbf{h}_{rq}&={\rm Fusion}(\mathbf{\hat e}^0_{c_0},\mathbf{\hat e}^1_{c_1},\ldots,\mathbf{\hat e}^{m-1}_{c_{m-1}}) \\ \hat y_{rq}&={\rm cos}(\mathbf{h}_{rq}, \mathbf{v}_{item})
\end{align}

\subsection{Target-Attention Net}

The Target-Attention Net acts as the offline teacher, generating high-fidelity supervision signals for the online student (the RQ-Attention Net). It computes the "ground-truth" interaction representation, $\mathbf{h}_{tar}$, by applying the full, computationally expensive TA mechanism to the user sequence $\mathbf{X}_{seq}$ and target item $\mathbf{x}_{item}$.
\begin{align}
\mathbf{H}^{tar}_{seq}&={\rm MHA}(\mathbf{X}_{seq}, \mathbf{X}_{seq}, \mathbf{X}_{seq}) \label{eq:selfAtt2},\\ 
\mathbf{h}_{tar}&={\rm MHA}(\mathbf{x}_{item},\mathbf{H}^{tar}_{seq},\mathbf{H}^{tar}_{seq})\label{eq:personalCB}.
\end{align}
This teacher representation $\mathbf{h}_{tar}$ then guides the training of the student's output $\mathbf{h}_{rq}$ via a knowledge distillation loss, where $sg$ ensure gradients only flow to the student:
\begin{align}
{\mathcal L}_{distill} = ||{\rm sg}[\mathbf{h}_{tar}]-\mathbf{h}_{rq}||_2^2\label{eq:distillLoss}.
\end{align}
For joint optimization, the teacher path also produces a relevance score $\hat y_{tar}$:
\begin{align}
\hat y_{tar}={\rm cos}(\mathbf{h}_{tar},\mathbf{v}_{item})\label{eq:scoreTA}.
\end{align}

\subsection{Codebook Alignment}

Given $m$ codebooks $\{\mathbf{\hat C}_l\}_{l=0}^{m-1}$, when the process begins with the initial residual $\mathbf{\hat r}_0=\mathbf{h}_{tar}$, it will yield another list of semantic IDs $\mathbf{\hat s}=[\hat c_0, \hat c_1, \ldots, \hat c_{m-1}]$ and $\mathbf{\hat E}_{tar}=[\mathbf{\hat e}^0_{\hat c_0}, \mathbf{\hat e}^1_{\hat c_1}, \ldots, \mathbf{\hat e}^{m-1}_{\hat c_{m-1}}]$. Evidently, $\sum_{l=0}^{m-1}\mathbf{\hat e}_{\hat c_l}^l$ is a more accurate approximation of $\mathbf{h}_{tar}$. To improve the consistency between $\mathbf{s}$ and $\mathbf{\hat s}$, we introduce a level-wise alignment strategy between the original and personalized codebooks. Specifically, at each RQ level $l$, we remodel the choice of a codebook entry $\mathbf{e}^l_i \in \mathbf{C}_l$ for a given residual $\mathbf{r}_l$ as a probability distribution, formulated via a softmax over the squared Euclidean distances:
\begin{align}
P(i|\mathbf{r}_l)=\frac{{\rm exp}(-||\mathbf{r}_l-\mathbf{e}^l_i||^2)}{\sum_{k=1}^K{\rm exp}(-||\mathbf{r}_l-\mathbf{e}^l_k||^2)}\label{eq:probabilitySelect}.
\end{align}
In this way, we can obtain the semantic ID distribution $P^l_\mathbf{z}$ over the codebook $\mathbf{C}_l$ with the initial residual $\mathbf{r}_0=\mathbf{z}$. Similarly, we can also obtain the semantic ID distribution $\hat P^l_{\mathbf{h}_{tar}}$ over the codebook $\mathbf{\hat C}_l$ with the initial residual $\mathbf{\hat r}_0=\mathbf{h}_{tar}$. We then use the Kullback-Leibler divergence to constrain the similarity between these two distributions. The loss function is as follows:
\begin{align}
{\mathcal L}_{align}=-\sum_{l=0}^{m-1}(D_{KL}(P^l_\mathbf{z}||\hat P^l_{\mathbf{h}_{tar}}) + D_{KL}(\hat P^l_{\mathbf{h}_{tar}}||P^l_\mathbf{z}))\label{eq:alignLoss},
\end{align}
where $D_{KL}(\cdot)$ is the Kullback-Leibler divergence between two probablity distribution.

Moreover, another benefit of our alignment strategy is the mitigation of codebook collapse. This phenomenon, where only a small subset of codebook entries are actively utilized, is known to significantly impair model performance. A widely-adopted solution is to initialize the codebook using k-means on the first training batch \cite{zeghidour2021}, which essentially aligns the initial codebook distribution with that of its inputs. Re-examining our alignment loss, we observe a similar mechanism. The softmax function enables each entry in codebooks to interact with all items. The direction of this interaction—whether it pulls them closer or pushes them apart—is determined by the discrepancy between the two distributions, $P^l_\mathbf{z}$ and $\hat P^l_{\mathbf{h}_{tar}}$. This process encourages the codebook to be close to the global item distribution, thereby significantly alleviating codebook collapse and boosting code utilization.

\subsection{Training Objective}

The entire model is trained end-to-end for Click-Through Rate (CTR) prediction. The final prediction $\hat y$ is an ensemble, combining the logit from the two-tower backbone, $\hat y_{tt}$, with the approximated interaction logit from RQ-Attention Net (student), $\hat y_{rq}$. This is optimized using the standard binary cross-entropy (BCE) loss against the ground-truth labels $y \in \{0, 1\}$:
\begin{align}
\hat y&=\sigma(\hat y_{tt} + \hat y_{rq})\label{eq:finalPred},\\
{\mathcal L}_{ctr}&=-y(\hat y)-(1-y)(1-\hat y)\label{eq:ctrLoss}.
\end{align}
To ensure Target-Attention Net provides meaningful supervision, its own output $\hat y_{tar}$ is concurrently optimized via its own BCE loss:
\begin{align}
\hat y^\prime&=\sigma(\hat y_{tt} + \hat y_{tar})\label{eq:tarPred},\\
{\mathcal L}_{tar}&=-y(\hat y^\prime)-(1-y)(1-\hat y^\prime)\label{eq:tarCtrLoss}.
\end{align}
The final overall loss function is a weighted sum of all components:
\begin{align}
{\mathcal L}
=\lambda_{1}{\mathcal L}_{ctr}
+\lambda_{2}{\mathcal L}_{tar}
+\lambda_{3}{\mathcal L}_{recon}
+\lambda_{4}{\mathcal L}_{rq}
+\lambda_{5}{\mathcal L}_{distill}
+\lambda_{6}{\mathcal L}_{align}\label{eq:finalLoss},
\end{align}
where $\lambda_{1}$ through $\lambda_{6}$ are hyperparameters used to control the weight of each respective loss term. We determined their values using a step-wise tuning strategy on a held-out validation set. First, to establish a reference scale for the loss landscape, we fix $\lambda_{1}$ = 1.0. Since $\lambda_{2}$ governs a comparable loss term, we set it to the same value ($\lambda_{2}$ = 1.0) to ensure a balanced training focus. The parameter $\lambda_{3}$ controls an auxiliary task whose gradients do not propagate back to the main network. We found the model to be insensitive to this hyperparameter and thus set it to 1.0 for simplicity. The remaining sensitive parameters, $\lambda_{4}$, $\lambda_{5}$, and $\lambda_{6}$, which control the delicate balance of the quantization and alignment losses, were empirically tuned via a grid search. The optimal configuration was found to be \{0.1, 1.0, 0.8\}, respectively. This set of hyperparameters was used for all subsequent offline and online experiments.


\subsection{Online Efficiency}
TARQ is designed for highly efficient online inference. The architecture achieves this through a strategic separation of computation:\\
\noindent{\bfseries Offline Pre-computation}: The expensive Target-Attention Net is confined to the training phase. Concurrently, the Residual-Quantizer component in RQ-Attention Net generates and stores a compact set of semantic IDs for every item in the corpus.\\
\noindent{\bfseries Online Inference}: At serving time, computation is decoupled from the candidate set size. The user tower and the core RQ-Attention component in RQ-Attention Net are computed only once per request. The complexity of the online attention mechanism scales not with the number of candidate items, but only with the total number of entries in the codebooks.\\
In our configuration with 8 codebooks of 16 entries each, the online attention mechanism processes a mere $8 \times 16=128$ vectors, regardless of the candidate set size. This fixed workload is orders of magnitude smaller than a full TA that would need to process tens of thousands of items per request, making TARQ eminently suitable for low-latency industrial recommender systems.

\section{Experiments}
\subsection{Experimental Setup}

\noindent{\bfseries Dataset.} Our experiments are conducted on a large-scale industrial dataset collected from the logs of the Taobao "618" Shopping Carnival. We enforce a strict temporal split, using a 19-day period for training and the subsequent day for testing. Clicked items serve as positive samples. The negative set consists of both un-clicked impressions and ranked but unexposed items from the production system. The final dataset contains over 14.2 billion interactions from $\sim175$ million users and $\sim20$ million items.

\noindent{\bfseries Baselines and Implementation.} We benchmark TARQ against strong baselines: (1) Two-Tower (TT), our core production model; (2) IntTower \cite{li2022} and MVKE \cite{xu2022}, representative models that enhance the TT architecture with cross-tower interactions. For a fair comparison, all models are implemented within a unified framework, sharing the same feature set and optimization settings.

\begin{table}[htbp]
\vspace*{-20pt}
  \setlength{\abovecaptionskip}{0cm}
  \setlength{\belowcaptionskip}{0.5cm}
  \centering
  \caption{Offline performance (AUC) and ablation study.} 
  \label{tab1}
  \setlength{\tabcolsep}{1mm}{
  \begin{tabular}{cc}
    \toprule
    Method & AUC \\
    \midrule
    Two-Tower (TT)     & 0.785 \\
    IntTower          & 0.787 \\
    MVKE              & 0.789 \\
    \textbf{TARQ (Full Model)}              & \textbf{0.799} \\
    TARQ w/o Codebook Alignment & 0.795 \\
    TARQ w/o Codebook Alignment \& Target-Attention Net & 0.796 \\
    TAVQ & 0.792 \\
    \bottomrule
  \end{tabular}}
\vspace*{-10pt}
\end{table}

\subsection{Offline Performance Comparison}
As shown in Table \ref{tab1}, TARQ achieves a state-of-the-art AUC of 0.799, outperforming the strong TT baseline by a substantial 0.014.

Crucially, the consistent gains from IntTower and MVKE over the TT baseline validate our core hypothesis: incorporating fine-grained interactions is critical for pre-ranking performance. However, TARQ’s significant lead over these other cross-tower models underscores the superiority of our TA approximation architecture, which more effectively captures these complex signals.


\subsection{Ablation Study}
We conducted an ablation study (Table \ref{tab1}) to isolate the contribution of each key component. First, we observed that augmenting the TT baseline with just RQ-Attention Net yielded a significant AUC gain of 0.011 (0.785 -> 0.796). However, when we further introduced Target-Attention Net by itself (without Codebook Alignment), the AUC paradoxically dropped from 0.796 to 0.795. Performance was only recovered and substantially improved when Target-Attention Net and Codebook Alignment were jointly applied. This validates that the absence of Codebook Alignment is highly detrimental to the model's effectiveness. Moreover, Codebook Alignment also substantially increased the overall codebook utilization from 59\% to 98\%, significantly mitigating the issue of codebook collapse. To isolate the contribution of RQ, we conduct another ablation study by simplifying it to a standard VQ (a single codebook layer of 128 vectors), creating a variant named TAVQ. In this variant, the attention mechanism operates over cluster centroids. Despite an identical computational cost, TAVQ's AUC drops by 0.007, confirming the significance of the multi-level quantization provided by RQ.


\subsection{Online A/B Testing}
To validate its real-world impact, we conducted an 8-day online A/B test on the Taobao platform during the "618" peak traffic period (June 13-20, 2025), with each experimental group comprising approximately 500,000 daily users. Against a highly-optimized production TT baseline, our proposed model, TARQ, achieved statistically significant improvements across the entire conversion funnel: \textbf{+0.57\%} relative lift in CTR (p=0.0498), \textbf{+4.59\%} in Conversion Rate (CVR) (p<0.0001), and \textbf{+7.57\%} in Gross Merchandise Volume (GMV) (p=0.0356).



\noindent{\bfseries Discussion on CVR/GMV Gains.} Interestingly, despite being trained solely on a CTR objective, TARQ yields significant online improvements in CVR and GMV. We attribute this phenomenon to \textit{selection bias} inherent in the training data. Our production system's global objective is to maximize transaction value. Consequently, items surviving the ranking funnel for exposure are biased towards high predicted transaction value. In learning to separate this population from the lower-value unexposed items, the model implicitly captures the latent transaction signals that differentiate them, not just click propensity. This hypothesis is strongly supported by the substantial CVR and GMV gains observed in live A/B tests.

\section{Conclusion}
In this paper, we introduced TARQ, a novel framework that, for the first time, successfully brings the modeling power of TA to the latency-critical pre-ranking stage. The principles of our RQ-based approximation are generalizable. Applying it to accelerate other computationally expensive interaction models (e.g., in the matching stage) presents another exciting avenue for future research.

\begin{acks}
Thanks to the colleagues in Taobao \& Tmall Group of Alibaba. This research was supported by Xidian University Specially Funded Project for Interdisciplinary Exploration (TZJHF202506).
\end{acks}

\bibliographystyle{ACM-Reference-Format}
\balance
\bibliography{sample-base}

@String{Computing = "Computing" }

@String{Computer = "{IEEE} Computer" }

@article{vaswani2017attention,
  title={Attention is all you need},
  author={Vaswani, Ashish and Shazeer, Noam and Parmar, Niki and Uszkoreit, Jakob and Jones, Llion and Gomez, Aidan N and Kaiser, {\L}ukasz and Polosukhin, Illia},
  journal={Advances in neural information processing systems},
  volume={30},
  year={2017}
}

@inproceedings{lee2022autoregressive,
  title={Autoregressive image generation using residual quantization},
  author={Lee, Doyup and Kim, Chiheon and Kim, Saehoon and Cho, Minsu and Han, Wook-Shin},
  booktitle={Proceedings of the IEEE/CVF conference on computer vision and pattern recognition},
  pages={11523--11532},
  year={2022}
}

@inproceedings{wang2025scaling,
  title={Scaling transformers for discriminative recommendation via generative pretraining},
  author={Wang, Chunqi and Wu, Bingchao and Chen, Zheng and Shen, Lei and Wang, Bing and Zeng, Xiaoyi},
  booktitle={Proceedings of the 31st ACM SIGKDD Conference on Knowledge Discovery and Data Mining V. 2},
  pages={2893--2903},
  year={2025}
}

@inproceedings{zheng2024adapting,
  title={Adapting large language models by integrating collaborative semantics for recommendation},
  author={Zheng, Bowen and Hou, Yupeng and Lu, Hongyu and Chen, Yu and Zhao, Wayne Xin and Chen, Ming and Wen, Ji-Rong},
  booktitle={2024 IEEE 40th International Conference on Data Engineering (ICDE)},
  pages={1435--1448},
  year={2024},
  organization={IEEE}
}

@article{feng2022recommender,
  title={Recommender forest for efficient retrieval},
  author={Feng, Chao and Li, Wuchao and Lian, Defu and Liu, Zheng and Chen, Enhong},
  journal={Advances in Neural Information Processing Systems},
  volume={35},
  pages={38912--38924},
  year={2022}
}

@article{rajput2023recommender,
  title={Recommender systems with generative retrieval},
  author={Rajput, Shashank and Mehta, Nikhil and Singh, Anima and Hulikal Keshavan, Raghunandan and Vu, Trung and Heldt, Lukasz and Hong, Lichan and Tay, Yi and Tran, Vinh and Samost, Jonah and others},
  journal={Advances in Neural Information Processing Systems},
  volume={36},
  pages={10299--10315},
  year={2023}
}

@article{tay2022transformer,
  title={Transformer memory as a differentiable search index},
  author={Tay, Yi and Tran, Vinh and Dehghani, Mostafa and Ni, Jianmo and Bahri, Dara and Mehta, Harsh and Qin, Zhen and Hui, Kai and Zhao, Zhe and Gupta, Jai and others},
  journal={Advances in Neural Information Processing Systems},
  volume={35},
  pages={21831--21843},
  year={2022}
}

@inproceedings{si2024twin,
  title={Twin v2: Scaling ultra-long user behavior sequence modeling for enhanced ctr prediction at kuaishou},
  author={Si, Zihua and Guan, Lin and Sun, ZhongXiang and Zang, Xiaoxue and Lu, Jing and Hui, Yiqun and Cao, Xingchao and Yang, Zeyu and Zheng, Yichen and Leng, Dewei and others},
  booktitle={Proceedings of the 33rd ACM International Conference on Information and Knowledge Management},
  pages={4890--4897},
  year={2024}
}

@inproceedings{chang2023twin,
  title={TWIN: TWo-stage interest network for lifelong user behavior modeling in CTR prediction at kuaishou},
  author={Chang, Jianxin and Zhang, Chenbin and Fu, Zhiyi and Zang, Xiaoxue and Guan, Lin and Lu, Jing and Hui, Yiqun and Leng, Dewei and Niu, Yanan and Song, Yang and others},
  booktitle={Proceedings of the 29th ACM SIGKDD Conference on Knowledge Discovery and Data Mining},
  pages={3785--3794},
  year={2023}
}

@inproceedings{cao2022sampling,
  title={Sampling is all you need on modeling long-term user behaviors for CTR prediction},
  author={Cao, Yue and Zhou, Xiaojiang and Feng, Jiaqi and Huang, Peihao and Xiao, Yao and Chen, Dayao and Chen, Sheng},
  booktitle={Proceedings of the 31st ACM International Conference on Information \& Knowledge Management},
  pages={2974--2983},
  year={2022}
}

@article{chen2021end,
  title={End-to-end user behavior retrieval in click-through rateprediction model},
  author={Chen, Qiwei and Pei, Changhua and Lv, Shanshan and Li, Chao and Ge, Junfeng and Ou, Wenwu},
  journal={arXiv preprint arXiv:2108.04468},
  year={2021}
}

@inproceedings{pi2020search,
  title={Search-based user interest modeling with lifelong sequential behavior data for click-through rate prediction},
  author={Pi, Qi and Zhou, Guorui and Zhang, Yujing and Wang, Zhe and Ren, Lejian and Fan, Ying and Zhu, Xiaoqiang and Gai, Kun},
  booktitle={Proceedings of the 29th ACM International Conference on Information \& Knowledge Management},
  pages={2685--2692},
  year={2020}
}

@inproceedings{chen2019behavior,
  title={Behavior sequence transformer for e-commerce recommendation in alibaba},
  author={Chen, Qiwei and Zhao, Huan and Li, Wei and Huang, Pipei and Ou, Wenwu},
  booktitle={Proceedings of the 1st international workshop on deep learning practice for high-dimensional sparse data},
  pages={1--4},
  year={2019}
}

@article{feng2019deep,
  title={Deep session interest network for click-through rate prediction},
  author={Feng, Yufei and Lv, Fuyu and Shen, Weichen and Wang, Menghan and Sun, Fei and Zhu, Yu and Yang, Keping},
  journal={arXiv preprint arXiv:1905.06482},
  year={2019}
}

@inproceedings{zhou2019deep,
  title={Deep interest evolution network for click-through rate prediction},
  author={Zhou, Guorui and Mou, Na and Fan, Ying and Pi, Qi and Bian, Weijie and Zhou, Chang and Zhu, Xiaoqiang and Gai, Kun},
  booktitle={Proceedings of the AAAI conference on artificial intelligence},
  volume={33},
  number={01},
  pages={5941--5948},
  year={2019}
}

@inproceedings{zhou2018deep,
  title={Deep interest network for click-through rate prediction},
  author={Zhou, Guorui and Zhu, Xiaoqiang and Song, Chenru and Fan, Ying and Zhu, Han and Ma, Xiao and Yan, Yanghui and Jin, Junqi and Li, Han and Gai, Kun},
  booktitle={Proceedings of the 24th ACM SIGKDD international conference on knowledge discovery \& data mining},
  pages={1059--1068},
  year={2018}
}

@inproceedings{gallagher2019joint,
  title={Joint optimization of cascade ranking models},
  author={Gallagher, Luke and Chen, Ruey-Cheng and Blanco, Roi and Culpepper, J Shane},
  booktitle={Proceedings of the twelfth ACM international conference on web search and data mining},
  pages={15--23},
  year={2019}
}

@inproceedings{liu2017cascade,
  title={Cascade ranking for operational e-commerce search},
  author={Liu, Shichen and Xiao, Fei and Ou, Wenwu and Si, Luo},
  booktitle={Proceedings of the 23rd ACM SIGKDD International Conference on Knowledge Discovery and Data Mining},
  pages={1557--1565},
  year={2017}
}

@article{guo2020deep,
  title={A deep look into neural ranking models for information retrieval},
  author={Guo, Jiafeng and Fan, Yixing and Pang, Liang and Yang, Liu and Ai, Qingyao and Zamani, Hamed and Wu, Chen and Croft, W Bruce and Cheng, Xueqi},
  journal={Information Processing \& Management},
  volume={57},
  number={6},
  pages={102067},
  year={2020},
  publisher={Elsevier}
}

@ArtifactSoftware{R,
    title = {R: A Language and Environment for Statistical Computing},
    author = {{R Core Team}},
    organization = {R Foundation for Statistical Computing},
    address = {Vienna, Austria},
    year = {2019},
    url = {https://www.R-project.org/},
}

@inproceedings{li2022,
  title={Inttower: the next generation of two-tower model for pre-ranking system},
  author={Li, Xiangyang and Chen, Bo and Guo, HuiFeng and Li, Jingjie and Zhu, Chenxu and Long, Xiang and Li, Sujian and Wang, Yichao and Guo, Wei and Mao, Longxia and others},
  booktitle={Proceedings of the 31st ACM International Conference on Information \& Knowledge Management},
  pages={3292--3301},
  year={2022}
}

@inproceedings{xu2022,
  title={Mixture of virtual-kernel experts for multi-objective user profile modeling},
  author={Xu, Zhenhui and Zhao, Meng and Liu, Liqun and Xiao, Lei and Zhang, Xiaopeng and Zhang, Bifeng},
  booktitle={Proceedings of the 28th ACM SIGKDD Conference on Knowledge Discovery and Data Mining},
  pages={4257--4267},
  year={2022}
}

@article{zeghidour2021,
  title={Soundstream: An end-to-end neural audio codec},
  author={Zeghidour, Neil and Luebs, Alejandro and Omran, Ahmed and Skoglund, Jan and Tagliasacchi, Marco},
  journal={IEEE/ACM Transactions on Audio, Speech, and Language Processing},
  volume={30},
  pages={495--507},
  year={2021},
  publisher={IEEE}
}










\end{document}